\documentclass[aps,prl,reprint,twocolumn,superscriptaddress,showpacs]{revtex4-1}
\usepackage{CJK}
\usepackage{graphicx}
\usepackage{mathrsfs}
\usepackage{bm}
\usepackage{amsmath}
\usepackage{dcolumn}
\usepackage{epstopdf}
\usepackage{dsfont}
\usepackage{amssymb}
\usepackage{tabularx}
\usepackage{array}
\usepackage{float}
\usepackage{color}
\usepackage{epstopdf}
\usepackage{mathrsfs}
\usepackage[colorlinks, linkcolor=blue,anchorcolor=blue,citecolor=blue,urlcolor=blue]{hyperref}

\begin{document}

\title{Composite Dirac Semimetal}

\author{Ziming Zhu}\email{These authors contributed equally to this work.}
\affiliation{Key Laboratory of Low-Dimensional Quantum Structures and Quantum Control of Ministry of Education, Department of Physics and Synergetic Innovation Center for Quantum Effects and Applications, Hunan Normal University, Changsha 410081, China}
\affiliation{Research Laboratory for Quantum Materials, Singapore University of Technology and Design, Singapore 487372, Singapore}

\author{Zhi-Ming Yu}\email{These authors contributed equally to this work.}
\affiliation{Research Laboratory for Quantum Materials, Singapore University of Technology and Design, Singapore 487372, Singapore}

\author{Weikang Wu}
\affiliation{Research Laboratory for Quantum Materials, Singapore University of Technology and Design, Singapore 487372, Singapore}

\author{Lifa Zhang}
\affiliation{Center for Quantum Transport and Thermal Energy Science, School of Physics and Technology, Nanjing Normal University, Nanjing 210023, China}

\author{Wei Zhang}
\affiliation{Fujian Provincial Key Laboratory of Quantum Manipulation and New Energy Materials, College of Physics and Energy, Fujian Normal University, Fuzhou 350117, China}

\author{Fan Zhang} \email{zhang@utdallas.edu}
\affiliation{Department of Physics, University of Texas at Dallas, Richardson, Texas 75080, USA}

\author{Shengyuan A. Yang}\email{shengyuan\_yang@sutd.edu.sg}
\affiliation{Research Laboratory for Quantum Materials, Singapore University of Technology and Design, Singapore 487372, Singapore}
\affiliation{Center for Quantum Transport and Thermal Energy Science, School of Physics and Technology, Nanjing Normal University, Nanjing 210023, China}

\begin{abstract}
Weak topological insulators and Dirac semimetals are gapped and nodal phases with distinct topological properties, respectively.
Here, we propose a novel topological phase that exhibits features of both and is dubbed composite Dirac semimetal (CDSM).
In its bulk, the CDSM has a pair of Dirac points and a pair of bands inverted along a high-symmetry path.
At side surfaces,  a pair of Fermi arcs connecting the projected Dirac points
coexist with a pair of Fermi loops traversing the surface Brillouin zone.
A nonsymmorphic symmetry dictates degeneracies between the Fermi arcs and the Fermi loops.
We characterize the CDSM by multiple topological invariants and show that, under a transition without breaking any symmetry,
it deforms into a topological crystalline insulator hosting two pairs of surface Fermi loops.
We demonstrate the CDSM in two models and predict its realization in the KAuTe-family materials.
\end{abstract}

\maketitle


The past decade has witnessed the predictive power of topological band theory and its applications to various materials.
In three dimensions (3D), a topological insulator is characterized by nontrivial $\mathbb{Z}_2$ invariants defined for its bulk band structure,
while at surfaces it features protected surface states~\cite{hasan2010colloquium,moore2010je,shen2012topological,fu2007topological,moore2007topological,roy2009topological}.
In a simple picture, the nontrivial band topology may be interpreted as an inverted band ordering
between the conduction and valence bands as compared with the atomic limit.
For example, if band inversion occurs only at the $\Gamma$ point, the resulting phase is a strong topological insulator
with one surface Dirac cones at every surface. On the other hand, if band inversion occurs along a high symmetry path,
it can give rise to a weak topological insulator (WTI) with two surface Dirac cones or Fermi loops only at those surfaces parallel to the path~\cite{fu2007topological,liu2016WTI}.

Band topology can also be used to characterize nodal phases such as semimetals~\cite{volovik2003universe,murakami2007phase,Wan2011,burkov2011weyl,young2012dirac,wang2012dirac,wang2013three,zhao2013topological,steinberg2014bulk,yang2014DWSC,yang2014classification,Chiu_RMP,burkov2016topological,yang2016sa,dai2016quantum,Armitage_RMP,Zhu2016,wu_Dirac,chen_Dirac,Zhu_Dirac,PhysRevX.7.041069,slager2013space}.
For example, in so-called Weyl semimetals~\cite{murakami2007phase,Wan2011,burkov2011weyl},
the conduction and valence bands cross linearly at twofold degenerate Weyl points; the low-energy quasiparticles resemble the Weyl fermions.
In the presence of both time reversal ($\mathcal{T}$) and inversion ($\mathcal{P}$) symmetry,
two Weyl points with opposite Chern numbers must merge into a fourfold degenerate Dirac point,
which can be further protected by crystalline symmetries~\cite{young2012dirac,steinberg2014bulk,wang2012dirac,wang2013three,chen2017ternary}.
The resulting phase is known as a Dirac semimetal,
in which the low-energy quasiparticles resemble the massless Dirac fermions. A Dirac superconductor was also predicted, with mirror symmetry-protected Dirac nodes~\cite{yang2014DWSC}.

The Dirac nodal phase may host a unique type of surface states:
Fermi arcs connecting the projections of the Dirac points on a surface~\cite{Wan2011,wang2012dirac,wang2013three,yang2014DWSC,wu_Dirac,chen_Dirac,Zhu_Dirac}.
Experimentally~\cite{liu2014discovery,borisenko2014experimental,liu2014stable,neupane2014observation},
the Dirac semimetals Na$_3$Bi and Cd$_3$As$_2$ have been demonstrated~\cite{wang2012dirac,wang2013three}, with a pair of Fermi arcs observed via the angle resolved photoemission spectroscopy (ARPES)~\cite{xu2015observation},
which can give rise to nonlocal cyclotron orbits~\cite{potter2014quantum,moll2016transport}. Thin films of these Dirac semimetals also offer a promising platform to achieve topological transistors~\cite{pan2015electric,guan2017artificial,collins2018electric}.

Since topological insulators and Dirac semimetals are characterized by different bulk topologies and different surface hallmarks, one may wonder {\it whether there exists a composite phase
that exhibits the features of both}. In this paper, we demonstrate the answer in the affirmative by establishing a theory that
highlights the first {\it composite Dirac semimetal} (CDSM),
which may be regarded as a stable combination of a WTI and a Dirac semimetal,
and show its realization in the KAuTe-family materials by using first-principles calculations.


\emph{\textcolor[rgb]{0.00,0.00,1.00}{Models and invariants.}}  We start by constructing an effective model which realizes the CDSM phase.
Consider a pair of Dirac points located on $\Gamma$-$A$ in the BZ,
similar to the situation of Dirac semimetals Na$_3$Bi and Cd$_3$As$_2$~\cite{wang2012dirac,wang2013three}.
Here, the $k_z$ axis (along $\Gamma$-$A$) is assumed to be the principal rotation axis,
which offers the symmetry protection needed for stabilizing the Dirac points~\cite{yang2014classification}.
We assume that the nontrivial band inversion is determined by the low-energy
bands along $\Gamma$-$A$, while the bands elsewhere are normally ordered and far away from Fermi energy.
Below, we establish an eight-band minimal model to capture the low-energy physics around the $\Gamma$-$A$ path.
This model has three distinctive features. (i) To stabilize the Dirac points,
the two crossing bands must belong to different irreducible representations of the symmetry group on $\Gamma$-$A$.
In our effective model, this indicates the decoupling of the two bands.
(ii) To exhibiting the WTI hallmarks,
there must be anther pair of bands that are completely inverted along $\Gamma$-$A$.
(iii) Each band must be Kramers degenerate in the presence of $\mathcal{T}$ and $\mathcal{P}$.

With these considerations, we construct the following eight-band model around the $\Gamma$-$A$ path
\begin{eqnarray}\label{Heff}
\mathcal{H}_\text{eff} & = & \left[\begin{array}{cc}
H_{\uparrow\uparrow} & \bm{0}\\
\bm{0} & H_{\downarrow\downarrow}
\end{array}\right],
\end{eqnarray}
where $H_{\uparrow\uparrow}$ and $H_{\downarrow\downarrow}$ are $4\times 4$ matrices with
\begin{eqnarray}
H_{\uparrow\uparrow}  =  \left[\begin{array}{cccc}
M_1 & B_{1}\cos\frac{k_{z}}{2} & 0 & Ak_{+}\\
B_{1}\cos\frac{k_{z}}{2} & M_1 & Ak_{+} & 0\\
0 & Ak_{-} & M_2 & B_2\cos\frac{k_{z}}{2}\\
Ak_{-} & 0 & B_2\cos\frac{k_{z}}{2} & M_2
\end{array}\right]
\end{eqnarray}
and $H_{\downarrow\downarrow}=H_{\uparrow\uparrow}^*$.
Here, $k_\pm=k_x\pm i k_y$ and $A$, $B_i$ and $M_i$ $(i=1,2)$ are model parameters.
On the $\Gamma$-$A$ path in which $k_\pm=0$, we obtain the following Kramers degenerate spectrum:
\begin{equation}
\varepsilon_{i,\pm}(k_z)=M_i\pm B_i\cos\frac{k_z}{2}.
\end{equation}
At $\Gamma$ and $A$, the band energies respectively read
\begin{equation}
\varepsilon_{i,\pm}^\Gamma=M_i\pm B_i,\quad \varepsilon_{i,\pm}^A=M_i.
\end{equation}
Note that the extra band degeneracy at $A$, not essential for the CDSM physics,
is dictated by a nonsymmorphic symmetry to be discussed later.

\begin{figure}[!htp]
{\includegraphics[clip,width=8.2cm]{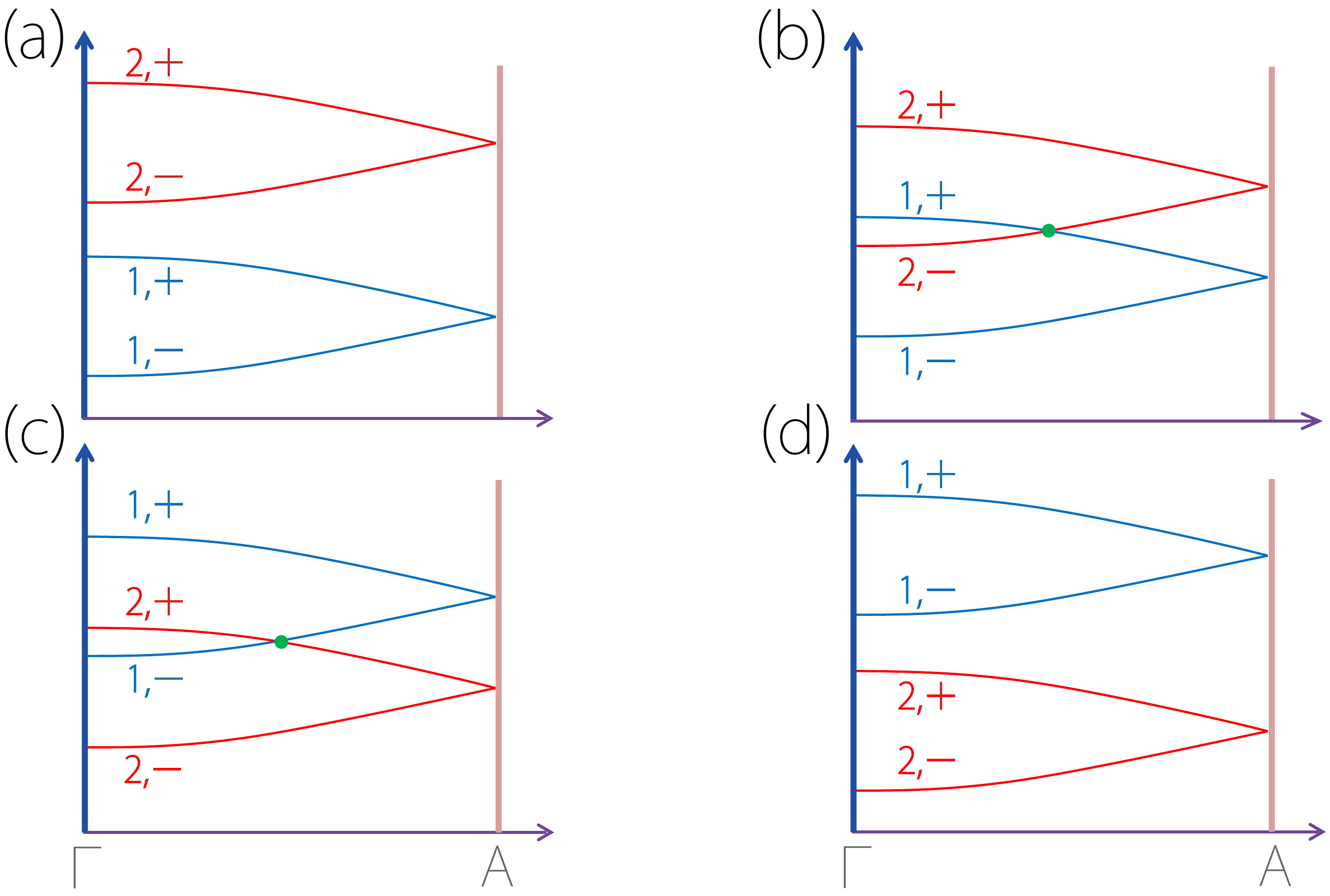}}
\caption{\label{fig:4_phases}
Four types of phases with distinct band ordering along $\Gamma$-$A$ path,
  as described by the effective model Eq.~(1). (c) is the CDSM phase. The green dots indicate the Dirac points.}
\end{figure}

By tuning $M_i$ and $B_i$, the band structures for model~(\ref{Heff}) can fall into four distinct phases, as illustrated in Fig.~\ref{fig:4_phases}.
Assume in the atomic limit the pair $\varepsilon_{2,\pm}$ is
energetically above $\varepsilon_{1,\pm}$, and consider the bands are half-filled.
The case in Fig.~\ref{fig:4_phases}(a) is a trivial insulating phase adiabatically connected to the atomic limit.
For the case in Fig.~\ref{fig:4_phases}(b),
there is only one band inversion at $\Gamma$, leading to the formation of Dirac points.
This is the usual Dirac semimetal phase.
When the $\varepsilon_{2,\pm}$ pair is further lowered in energy,
band inversion also occurs at $A$ as shown in Fig.~\ref{fig:4_phases}(c). In fact, at $A$ there is a double band inversion
since $\varepsilon_{1,\pm}^A>\varepsilon_{2,\pm}^A$, while at $\Gamma$ there is still a single band inversion. This also creates a pair of Dirac points,
while the other two bands $\varepsilon_{2,-}$ and $\varepsilon_{1,+}$ are completely inverted along $\Gamma$-$A$.
This is the CDSM phase, a key discovery of this work.
When the $\varepsilon_{2,\pm}$ pair is entirely below the $\varepsilon_{1,\pm}$ pair, as seen in Fig.~\ref{fig:4_phases}(d),
there is another insulating phase. Compared to the trivial one in Fig.~\ref{fig:4_phases}(a), this phase has a double band inversion
at both $\Gamma$ and $A$ and exhibit nontrivial band topology that to be characterized.
Hereafter, we will focus on the CDSM phase in Fig.~\ref{fig:4_phases}(c) and examine its possible transition to the topological phase in Fig.~\ref{fig:4_phases}(d).

\begin{figure}[!htp]
{\includegraphics[clip,width=8.2cm]{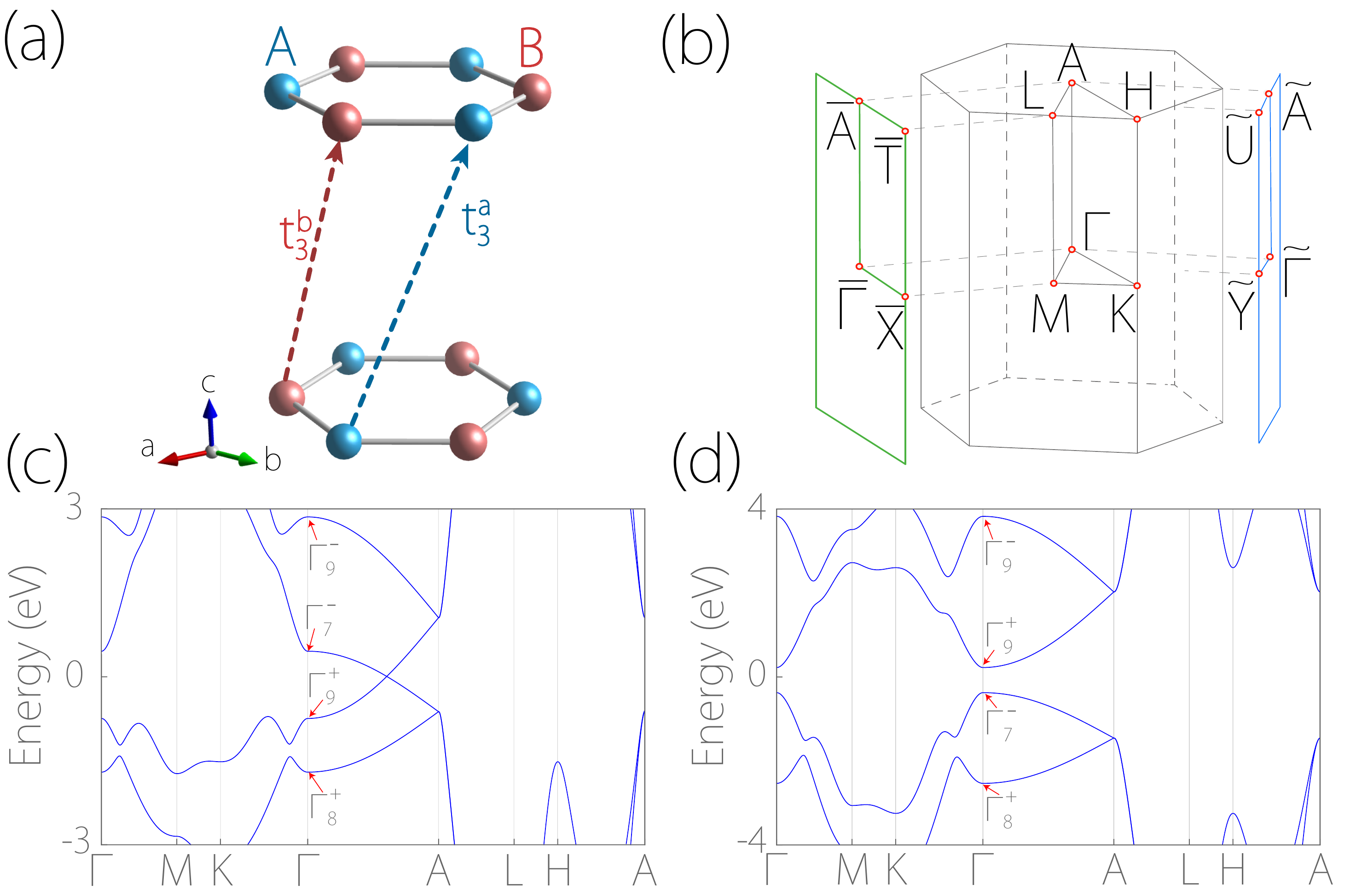}}
\caption{\label{fig:atomic_BZ}
(a) 3D lattice model. The arrows indicate the two interlayer hopping processes.
  (b) Bulk BZ, (100) surface BZ (green line), and (110) surface BZ (blue line).
  The corresponding crystal planes in Miller notation are respectively refereed as $(11\overline{2}0)$ and $(\overline{1}100)$.
  (c) Band structure of a CDSM and (d) that of a topological crystalline insulator obtained
  in the tight-binding model Eq.~(5), we have used $\epsilon_a=-1.2501$~eV,
  $\epsilon_b=0.5775$~eV, $t_1=-0.2842$~eV, $t_2^b=-0.3005$~eV, $t_3^a=0.2001$~eV,
  $t_3^b=-0.1202$~eV, and $t_2^a=0.1906$~eV in (c) and $0.4573$~eV in (d).}
\end{figure}

To fully characterize the CDSM phase and its nontrivial surface states, we extend the continuum
model~(\ref{Heff}) to a tight-binding model. Inspired by the concrete material to be discussed later,
we consider a 3D lattice consisting of 2D honeycomb layers stacked along $z$, as sketched in Fig.~\ref{fig:atomic_BZ}(a).
For each layer, the $A$ and $B$ sites are occupied by two different types of atoms, $a$ and $b$ respectively.
Each unit cell contains two layers, between which $a$ and $b$ are switched. We assume that each site has two basis orbitals forming a Kramers pair:
$|p_+,\uparrow\rangle$ and $|p_-,\downarrow\rangle$ on $a$,
whereas $|d_{+2},\uparrow\rangle$ and $|d_{-2},\downarrow\rangle$ on $b$,
where $p_\pm=p_x\pm ip_y$ and {$d_{\pm2}=d_{x^2-y^2}\pm 2i d_{xy}$}.
Based on these, we construct the following tight-binding model
\begin{eqnarray}\label{HTB}
\mathcal{H}&=&\sum_{\alpha,i}(\epsilon_a a^\dagger_{\alpha,i}a_{\alpha,i}+\epsilon_b b^\dagger_{\alpha,i}b_{\alpha,i})\nonumber\\
&+&\sum_{\alpha,i,m}t_1(-1)^\alpha(a^\dagger_{\alpha,i+\bm R_m}\sigma_z e^{i\frac{(2m-1)\pi}{3}\sigma_z}b_{\alpha,i}+h.c.)\nonumber\\
&+&\sum_{\alpha,i,n}(t_2^a a^\dagger_{\alpha,i+\bm R_n'}a_{\alpha,i}+t_2^b b^\dagger_{\alpha,i+\bm R_n'}b_{\alpha,i})\nonumber\\
&+&\sum_{i,m}(t_3^a a_{0,i+\bm R_m}^\dagger a_{1,i}+t_3^b b_{1,i+\bm R_m}^\dagger b_{0,i}+h.c.).
\end{eqnarray}
Here, $a^\dagger=(a^\dagger_{|p_+,\uparrow\rangle},a^\dagger_{|p_-,\downarrow\rangle})$
and $b^\dagger=(b^\dagger_{|d_{+2},\uparrow\rangle},b^\dagger_{|d_{-2},\downarrow\rangle})$
are the electron creation operators, $\alpha=0,1$ label the two layers in a unit cell,
$i$ labels the sites within a layer, $\bm R_m$ ($m=1,2,3$) correspond to the vectors connecting to
the three nearest neighbors in a layer, $\bm R_n'$ ($n=1,\cdots,6$) correspond to the vectors connecting to
the six next nearest neighbors in a layer, $\epsilon_a$ and $\epsilon_b$ are the on-site energies,
and the $t$'s are various hopping amplitudes (taken to be real).
In model~(\ref{HTB}), the first term represents an on-site energy difference,
and the second and third terms are hoppings within a honeycomb layer,
with the extra phase factor due to the different orbital characters on $a$ and $b$.
In this model, the nearest interlayer hopping is suppressed as the two orbitals involved
have different angular momenta along $z$.
Thus, the strongest interlayer hopping, i.e., the last term in model~(\ref{HTB}),
occurs between two $a$ or two $b$ sites, as indicated in Fig.~\ref{fig:atomic_BZ}(a).

The model has three important symmetries in addition to the $\mathcal{T}$ and $\mathcal{P}$:
a six-fold screw rotation $\widetilde{C}_6:\,(x,y,z)\rightarrow(x/2-\sqrt{3}y/2,\sqrt{3}x/2+y/2,z+1/2)$,
a horizontal mirror
 ${M_{z}}:\,(x,y,z)\rightarrow (x,y,-z+1/2)$,
and {a vertical glide mirror $\widetilde{M}_y:\,(x,y,z)\rightarrow (x,-y,z+1/2)$.}
These symmetries correspond to the space group $P6_{3}/mmc$.
It is straightforward to show that the tight-binding model~(\ref{HTB}) reduces to the effective model~(\ref{Heff})
around the $\Gamma$-$A$ path, with the identification of {$M_{1(2)}=\epsilon_{a(b)}+6t_{2}^{a(b)}$, $A=\sqrt{3} t_1$,
and $B_{1(2)}=6t_3^{a(b)}$}.

Fig.~\ref{fig:atomic_BZ}(c) shows the band structure of the CDSM phase in the tight-binding model~(\ref{HTB}),
and the low-energy physics along $\Gamma$-$A$ resembles that in Fig.~\ref{fig:4_phases}(c).
Here, the two bands corresponding to $\varepsilon_{1,\pm}$ have the $p$-orbital character,
whereas the other two corresponding to $\varepsilon_{2,\pm}$ have the $d$-orbital character.
In the atomic limit, the $d$-orbital has a higher energy than the $p$-orbital,
and the band structure in Fig.~\ref{fig:atomic_BZ}(c) satisfies the band inversion pattern required for the CDSM phase.
In addition, the two crossing bands on $\Gamma$-$A$ belong to different irreducible representations ($\Gamma_7$
and $\Gamma_9$) of $C_{6v}$ symmetry. Therefore, the Dirac point is symmetry-protected.

To further characterize the topology of CDSM phase, we examine possible bulk topological invariants.
Because a Dirac node exists between $\Gamma$ and $A$ points, and because the two planes $k_z=0$
and $k_z=\pi$ are fully gapped, the 2D topological invariants of the two planes must be topologically distinct.
We can consider their 2D $\mathbb{Z}_2$ invariants~\cite{kane2005z},
$\nu_{0}$ and $\nu_{\pi}$. Following the Fu-Kane method~\cite{fu2007topological_p},
we find that $(\nu_{0},\nu_{\pi})=(1,0)$. These two invariants do capture the single band inversion
at $\Gamma$ but cannot identify the double band inversion at $A$. Since the $M_z$ symmetry is also present,
we can further consider the mirror Chern numbers~\cite{teo2008surface,hsieh2012topological}
$\mathcal{N}_{0}$ and $\mathcal{N}_{\pi}$ for the $k_z=0$ and $k_z=\pi$ planes.
We find that $(\mathcal{N}_{0},\mathcal{N}_{\pi})=(1,2)$,
which respectively capture the single and double band inversion at $\Gamma$ and $A$.
We note that the usual DSM phase can be indexed by $(\mathcal{N}_{0},\mathcal{N}_{\pi})=(0,1)$ or $(1,0)$.
The established bulk invariants crucially determine the presence and connectivity of the surface states to be studied below.

\emph{\textcolor[rgb]{0.00,0.00,1.00}{Topological phase transition.}}  Indicated by the topological invariants, the breaking of the rotational symmetry
can gap the Dirac points and produce a strong topological insulator.
Here we are more interested in the topological phase transition during which all the symmetries are preserved,
i.e., the transition from the CDSM phase in Fig.~\ref{fig:atomic_BZ}(c) to an unusual topological insulator phase in Fig.~\ref{fig:atomic_BZ}(d).
This transition is associated with a band inversion process at $\Gamma$.
Consequently, the two Dirac points gradually move to and eventually annihilate at $\Gamma$,
such that double band inversion occurs at both $\Gamma$ and $A$.
For this insulating phase, we find that the 2D $\mathbb{Z}_2$ invariants are $(\nu_{0},\nu_{\pi})=(0,0)$,
and that the mirror Chern numbers are $(\mathcal{N}_{0},\mathcal{N}_{\pi})=(2,2)$.
To the best of our knowledge, such a topological crystalline insulator has not been studied before.
Below, we will show that this topological phase may be regarded as two copies of WTI,
with four protected surface Fermi loops.

\emph{\textcolor[rgb]{0.00,0.00,1.00}{Surface states.}}  The hallmark of CDSM phase is manifested by its exotic surface states.
Here we examine the surface spectrum based on the tight-binding model~(\ref{HTB}).
On the (001) surface, the two bulk Dirac points are projected to the same point,
and thus there should be no surface Fermi arcs. However, we note that there is still one helical surface state,
since its 3D $\mathbb{Z}_2$ invariants are $(1;000)$~\cite{fu2007topological_p},
which can be deduced from its 2D $\mathbb{Z}_2$ invariants $(\nu_{0},\nu_{\pi})=(1,0)$.

\begin{figure}[!htp]
{\includegraphics[clip,width=8.2cm]{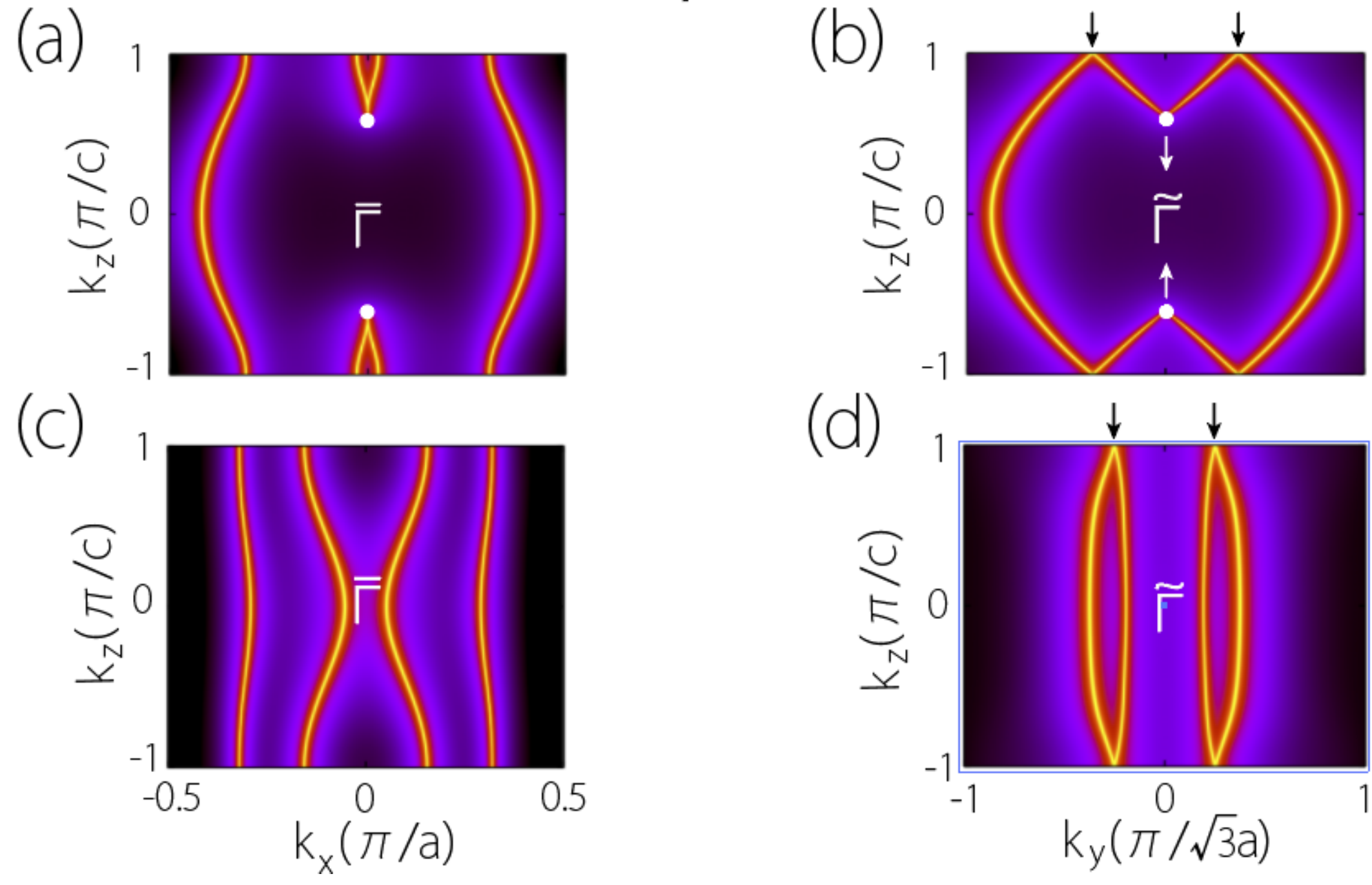}}
\caption{\label{fig:arcs_tb1}
Equal-energy contours at the Dirac point energy featuring the evolution of the surface states
  as the phase undergoes the transition from (a,b) CDSM to (c,d) topological crystalline insulator,
  with (a,c) for the (100) surface and (b,d) for the (110) surface.
  The model parameter values in (a)-(d) are the same to those used in Fig.~\ref{fig:atomic_BZ}.}
\end{figure}

It turns out that more interesting physics occurs at side surfaces. Consider the (100) surface first.
As featured in Fig.~\ref{fig:arcs_tb1}(a), the two bulk Dirac points are projected to the two sides of the $\overline{\Gamma}$ point, and they are connected by a pair of Fermi arcs.
Unlike the usual Dirac semimetals Na$_3$Bi and Cd$_3$As$_2$ for which
the Fermi arcs go around the $\overline{\Gamma}$ point~\cite{wang2012dirac,wang2013three},
the Fermi arcs here go around the $\overline{A}$ point and cross the surface BZ boundary.
Additionally, there is a pair of Fermi loops traversing the surface BZ, like those for a WTI~\cite{liu2016WTI}.
The emergence of the WTI-like surface states here can be attributed to the band inversion between
the two higher-energy bands long $\Gamma$-$A$ in the bulk,
corresponding to $\varepsilon_{1,+}$ and $\varepsilon_{2,-}$ in Fig.~\ref{fig:4_phases}(c).
Such a composite surface state pattern is required by the established bulk topological invariants.
Given that $\mathcal{N}_{0}=1$, the $k_z=0$ plane contributes one pair of ``edge'' states.
As a result, at the constant energy slice in Fig.~\ref{fig:arcs_tb1}(a), a pair of surface states appears in the $k_z=0$ line.
Similarly, given that $\mathcal{N}_{\pi}=2$, there must be two pairs of surface states in the $k_z=\pi$ line.

At the (110) surface, a composite pattern similar to the (100) surface should also appear.
However, there is one important difference. As featured in Fig.~\ref{fig:arcs_tb1}(b), at the (110) surface the Fermi arcs and Fermi loops
are connected at $k_z=\pi$, whereas such degeneracies are absent on the (100) surface.
These degeneracies are dictated by the aforementioned nonsymmorphic
glide mirror $\widetilde{M}_y$. The (110) surface preserves this symmetry.
Note that the composite anti-unitary symmetry $\mathcal{T}\widetilde{M}_y$ satisfies
\begin{equation}
  (\mathcal{T}\widetilde{M}_y)^2=e^{-ik_z}.
\end{equation}
On the $k_z=\pi$ line, because each point is invariant under $\mathcal{T}\widetilde{M}_y$
and because $(\mathcal{T}\widetilde{M}_y)^2=-1$, the surface states must form Kramers-like pairs.
This explains the origin of the surface state connectivity in Fig.~\ref{fig:arcs_tb1}(b).
By contrast, this double degeneracy is absent in Fig.~\ref{fig:arcs_tb1}(a),
since the (100) surface breaks the $\mathcal{T}\widetilde{M}_y$ symmetry.

Under the phase transition from the CDSM to the topological crystalline insulator, the surface states also transform [see Fig.~\ref{fig:arcs_tb1}(c)~and~3(d)]. As the two Dirac points move toward the $\Gamma$ point, the two Fermi arcs are elongated.
After they merge and annihilate, each arc transforms into a Fermi loop traversing the surface BZ.
Hence, the topological crystalline insulator phase may be regarded as two copies of WTI~\cite{liu2016WTI},
with four surface Fermi loops. This is indeed consistent with the bulk invariants
$(\mathcal{N}_{0},\mathcal{N}_{\pi})=(2,2)$. The same physics occurs at both the (100) and (110) surfaces,
except that in the latter case the $\mathcal{T}\widetilde{M}_y$ symmetry maintains the Kramers-like degeneracy
on the $k_z=\pi$ line.

\begin{figure}[!htp]
{\includegraphics[clip,width=8.2cm]{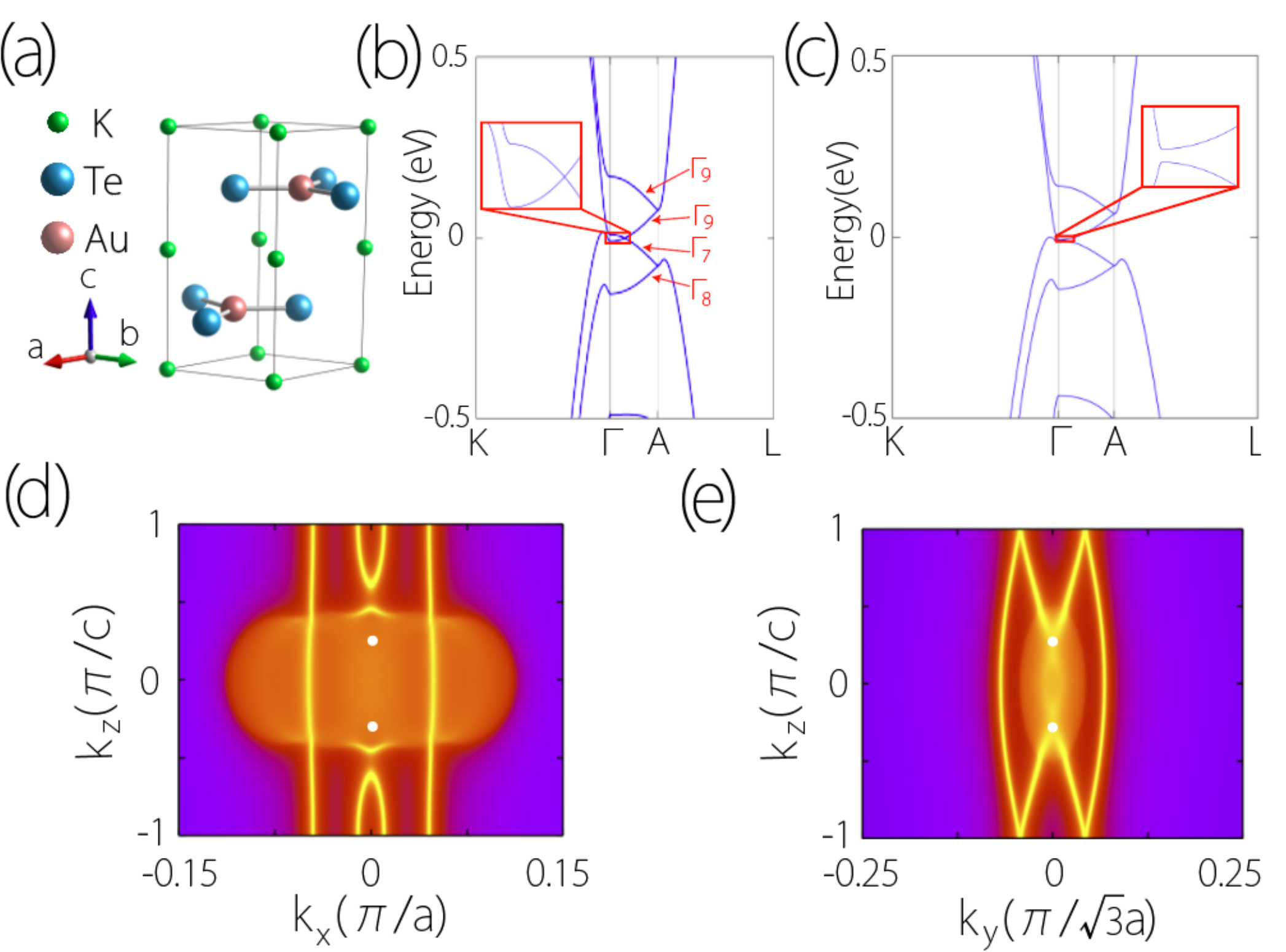}}
\caption{\label{fig:bands_pdos}
(a) Primitive unit cell of KAuTe. DFT results for the bulk band structure (SOC included) of (b) KAuTe and
  (c) RbAuTe. DFT results for KAuTe showing the equal-energy contour of (e) the (100) surface spectrum
  and (f) the (110) surface spectrum.}
\end{figure}


\emph{\textcolor[rgb]{0.00,0.00,1.00}{Material realization.}}  The physics of CDSM can be realized in the KAuTe-family materials.
Experimentally, the KAuTe single crystal was successfully by a fusion reaction of the elements at 823 K
and demonstrated to be stable at room temperature~\cite{bronger1990kaute,bronger1992synthese}.
As shown in Fig.~\ref{fig:bands_pdos}(a), KAuTe has a layered structure with the $P6_{3}/mmc$ space group symmetry.
Te and Au form planar honeycomb layers stacked along the $c$-axis.
Acting as charge donors, the K atoms are intercalated between adjacent Te-Au layers. If the K atoms were removed,
the crystal lattice would become identical to the lattice for our tight-binding model.

Fig.~\ref{fig:bands_pdos}(b) presents the band structure of KAuTe near the Fermi level, obtained by our first-principles calculations [see Supplemental Material].
The material shows the character of a semimetal~\cite{PhysRevLett.106.156402}.
On the $\Gamma$-$A$ path, the band features are similar to that in Fig.~\ref{fig:atomic_BZ}(c).
Around $A$, the two higher bands are dominated by Te $p$ orbitals
whereas the two lower bands are mainly from the Au $d$ orbitals.
Evidently, a double band inversion occurs at $A$ while a single at $\Gamma$, consistent with the scenario in Fig.~\ref{fig:atomic_BZ}(c).
As a result, the crossing between the $\Gamma_7$ and $\Gamma_9$ bands
is a symmetry-protected Dirac point, and the material is a CDSM.
We further evaluate the topological invariants of KAuTe based on our first-principles calculations
and find that indeed $(\mathcal{N}_{0},\mathcal{N}_{\pi})=(1,2)$ and $(\nu_{0},\nu_{\pi})=(1,0)$.
This unambiguously confirms that KAuTe is a CDSM.

Fig.~\ref{fig:bands_pdos}(d)~and~4(e) plot the (100) and (110) surface spectra of KAuTe.
Their similarity to Fig.~\ref{fig:arcs_tb1}(a,b) are clear, although there are additional projected bulk states.
Evidently, KAuTe exhibits the composite surface states, which are the hallmark of CDSM.
We note that by replacing K with heavier elements in the same group, such as Rb or Cs,
both the spin-orbit coupling and band inversion can be enhanced.
We find that RbAuTe has an extra band inversion at $\Gamma$ as shown in Fig.~\ref{fig:bands_pdos}(c)
and realizes the topological crystalline insulator phase with $(\mathcal{N}_{0},\mathcal{N}_{\pi})=(2,2)$.
This suggests that the transition from the CDSM phase to the topological crystalline phase
can also be realized by lattice deformation, e.g., via uniaxial strain along the $c$-axis.

Experimentally, the bulk and surface band structures for CDSM
can be directly probed via the ARPES~\cite{liu2014discovery,xu2015discovery}.
The surface states and their deformation under the topological phase transition can also be detected by surface sensitive probes, such as the scanning tunneling spectroscopy/microscopy ~\cite{zhang2009experimental,roushan2009topological}.
Particularly, the unique surface-state Fermi surface, with coexisting Fermi arcs and Fermi loops, may produce salient features
in the quasiparticle interference pattern~\cite{zheng2016atomic, inoue2016quasiparticle, zheng2018quasiparticle,lau2017generic}.

\bibliography{CDSM_refs}

\end{document}